\documentclass[12pt]{article}
\usepackage{graphicx} 
\usepackage[margin=2.7cm]{geometry}

\newcommand{\be}{\begin{equation}}
\newcommand{\ee}{\end{equation}}
\newcommand{\af}{{ \lambda }} 
\newcommand{\qhat}{ {\hat Q} } 
\newcommand{\mfont}{ \mathcal }

\newcommand{\sint}{{ {\rm Si} }}
\newcommand{\aint}{{ {\Gamma}}}
\newcommand{\eigenvalue}{{ {\Lambda} }}
\newcommand{\eangle}{{ \psi }} 
\newcommand{\period}{{ \Delta \tau }} 
\newcommand{\square}{ \bullet} 

\begin{document} 

$\,$ 
\vskip 1.0truein

\centerline{\bf HILL'S EQUATION WITH RANDOM FORCING PARAMETERS: } 
\medskip 
\centerline{\bf THE LIMIT OF DELTA FUNCTION BARRIERS} 

\bigskip 
\bigskip

\centerline{Fred C. Adams$^{1,2}$ and Anthony M. Bloch$^{1,3}$} 
\bigskip
\centerline{\it $^1$Michigan Center for Theoretical Physics, University of Michigan, Ann Arbor, MI 48109} 
\bigskip
\centerline{\it $^2$Astronomy Department, University of Michigan, Ann Arbor, MI 48109}
\bigskip
\centerline{\it $^3$Department of Mathematics, University of Michigan, Ann Arbor, MI 48109}
\bigskip

\begin{abstract} 

This paper considers random Hill's equations in the limit where the
periodic forcing function becomes a Dirac delta function. For this
class of equations, the forcing strength $q_k$, the oscillation
frequency $\af_k$, and the period $(\period)_k$ are allowed to vary
from cycle to cycle.  Such equations arise in astrophysical orbital
problems in extended mass distributions, in the reheating problem for
inflationary cosmologies, and in periodic Schr{\"o}dinger equations.
The growth rates for solutions to the periodic differential equation
can be described by a matrix transformation, where the matrix elements
vary from cycle to cycle. Working in the delta function limit, this
paper addresses several coupled issues: We find the growth rates for
the $2 \times 2$ matrices that describe the solutions. This analysis
is carried out in the limiting regimes of both large $q_k \gg 1$ and
small $q_k \ll 1$ forcing strength parameters.  For the latter case,
we present an alternate treatment of the dynamics in terms of a
Fokker-Planck equation, which allows for a comparison of the two
approaches.  Finally, we elucidate the relationship between the
fundamental parameters $(\af_k,q_k)$ appearing in the stochastic
differential equation and the matrix elements that specify the
corresponding discrete map. This work provides analytic --- and
accurate --- expressions for the growth rates of these stochastic
differential equations in both the $q_k \gg1 $ and the $q_k \ll 1$
limits.

\end{abstract} 

\newpage 
\baselineskip=18pt

\bigskip
\noindent
{\bf I. INTRODUCTION} 
\bigskip 

This paper considers random Hill's equations in the delta function
limit.  A random Hill's equation can be written in the form
\be 
{d^2 y \over dt^2} + [ \af_k + q_k \qhat (t) ] y = 0 \, , 
\label{basic} 
\ee 
where the barrier shape function $\qhat(t)$ is periodic, so that
$\qhat (t + \period) = \qhat(t)$, where $\period$ is the period (we
generally take $\period = \pi$). In the delta function limit, 
the periodic functions $\qhat(t)$ become Dirac delta functions, 
\be
\qhat(t) = \delta ([t]-\pi/2) \, , 
\label{qhatdelta} 
\ee
where the square brackets indicate that time is measured mod-$\pi$.
The parameter $q_k$ denotes the forcing strength, which is a random
variable that takes on a new value every cycle (where the index $k$
determines the cycle). The parameter $\af_k$, which determines the
natural oscillation frequency of the system, also varies from cycle to
cycle. In this work the period $\period$ is considered fixed; one can
show that cycle to cycle variations in $\period$ can be scaled out of
the problem and included in the distributions of the $(\af_k,q_k)$ [2].

The original form of Hill's equation (\ref{basic}) holds the values of
the parameters constant [13], and such equations arise often in
physics [19]. A straightforward generalization of this classic problem
is to consider parameters that vary from cycle to cycle (according to
a well-defined distribution). Further, the limit of delta function
barriers (equation [\ref{qhatdelta}]) arises in many applications (see
below) and thus provides a natural starting point for this analysis.

One specific motivation for considering random Hill's equations arises
from orbit problems in astrophysical settings, including dark matter
halos, galactic bulges, tidal streams, and young embedded star
clusters.  These astrophysical systems generally have non-spherical,
extended mass distributions, with corresponding potentials that are
asymmetric.  With this loss of symmetry, angular momentum is not
conserved, orbits are not confined to particular planes, and orbital
instabilities often arise. For example, if an orbit is initially
confined to the principal plane of a dark matter halo (or any
triaxial, extended mass distribution), the motion is unstable to
perturbations out of the orbital plane (see Ref. [4] and Appendix A).
The development of this instability [2] is described by a random
Hill's equation (as given by equation [\ref{basic}]), with
sharply-peaked forcing barriers that can be described by delta
functions (as given by equation [\ref{qhatdelta}]).  This orbit
instability arises in many other astrophysical systems, including
embedded young star clusters, galactic bulges, and tidal streams
[4,6]. This instability produces a number of astrophysical effects,
including changing the velocity distributions from radial to more
isotropic, making highly flattened systems more rounded, and helping
to disperse tidal streams.

In another application, the reheating epoch at the end of the
inflationary phase in the early universe [12] is described by a
``parametric resonance instability'' [17].  During inflation, the
potential of the inflaton field $\Phi$ dominates the energy density,
which is primarily in the form of vacuum energy.  After sufficient
inflation has taken place, this energy must be converted into matter
and radiation so that the universe can evolve into its present
state. As a result, the inflaton field $\Phi$ must couple to matter
fields $\chi$, and the subsequent conversion of energy is governed by
a Hill's equation, often the Mathieu equation [15]. Additional
fluctuations [15,20] that are present during this process --- due to
thermal and quantum effects --- convert the equation of motion for
reheating into a random Hill's equation (see Appendix B). During
inflationary reheating, the fluctuations have small amplitudes and the
forcing terms can be modeled as delta functions, so that the resulting
problem is described by equations (\ref{basic}) and
(\ref{qhatdelta}). The instability itself acts to rapidly convert the
vacuum energy of the universe into matter and radiation. In the
classical problem, however, the parameter space of Hill's equation
retains bands of stability that can inhibit this conversion. Random
fluctuations tend to erase these bands of stability, as shown herein,
and thereby increase the efficacy of the reheating process.  

For completeness, we note that in quantum systems with periodic
lattices and a source of noise, the corresponding Schr{\"o}dinger
equation takes the form of a random Hill's equation [5,7]. This topic
is relatively well developed [20], but the results presented in this
paper can also be useful in this context. 

Periodic differential equations in this class can be described by a
discrete mapping of the coefficients of the principal solutions from
one cycle to the next.  The transformation matrix takes the form
\be
{\mfont M}_k = \left[ 
\matrix{h_k & (h_k^2 - 1)/g_k \cr g_k & h_k} \right] \, , 
\label{mapzero} 
\ee
where the subscript denotes the cycle. The matrix elements for the 
$kth$ cycle are given by  
\be
h_k = y_1 (\pi) \qquad {\rm and} \qquad g_k = {\dot y}_1 (\pi) \, , 
\label{hk}
\ee
where $y_1$ and $y_2$ are the principal solutions for that cycle. For
Hill's equation with delta function barriers (\ref{qhatdelta}), the
principal solutions have been found previously [4] and the matrix 
elements take the form 
\be
h_k = \cos \varphi_k - {q_k \over 2 \sqrt{\af_k} } \sin\varphi_k 
\qquad {\rm and} \qquad 
g_k = - \sqrt{\af_k} \sin\varphi_k - q_k \cos^2 (\varphi_k/2) \, ,  
\label{deltaprin} 
\ee
where we have defined $\varphi_k = \sqrt{\af_k} \pi$. The index 
$k$ indicates that the quantities $(\af_k, q_k)$, and hence 
the solutions $(h_k, g_k)$, vary from cycle to cycle. Throughout 
this work, the random variables are taken to be independent and 
identically distributed (iid).  

Note that one might expect the matrix in equation $(\ref{mapzero})$ to
have four independent elements (instead of only two). In this paper,
however, we specialize to the case where the periodic functions
$\qhat(t)$ are delta functions, which are symmetric about the midpoint
of the period.  This property implies that $y_1(\pi) = {\dot y}_2
(\pi)$, which eliminates one independent matrix element [1,19]. In
addition, since the Wronskian of the original differential equation
(\ref{basic}) is unity, the determinant of the matrix map must be
unity, and this constraint eliminates another independent element.

The growth rates for Hill's equation (\ref{basic}) are determined by
the growth rates for matrix multiplication of the matrices 
${\mfont M}_k$ given by equation (\ref{mapzero}).  Here we denote the
product of $N$ such matrices as ${\mfont M}^{(N)}$, and the growth
rate $\gamma$ is defined by
\be
\gamma = \lim_{N \to \infty} {1 \over N} 
\log || {\mfont M}^{(N)} || \, . 
\label{growbasic}
\ee
Previous work [10,11,18] shows that this result is independent of the
choice of the norm $|| \cdot ||$.

This paper is organized as follows. To determine the growth rates for
the differential equation (\ref{basic}), we consider the
multiplication of infinite strings of random matrices of the form
(\ref{mapzero}). Working in the delta function limit, this paper
explores the regime where the forcing strengths are large $q_k \gg 1$
(Section II), and the opposite regime of small forcing parameters $q_k
\ll 1$ (Section III).  For the small $q_k$ limit, we develop an
alternate treatment of the dynamics using the Fokker-Planck equation
(Section IV).  Next, we consider the relationship between the matrix
elements $(h_k,g_k)$ that appear in the discrete map (\ref{mapzero})
and the random variables $(\af_k, q_k)$ that appear in the original
differential equation (\ref{basic}). For the limiting case of delta
functions barriers, we find this transformation explicitly (see
equation [\ref{deltaprin}]), and constrain the distributions of the
matrix elements for given distributions of the input parameters in 
Section V.  The paper concludes (in Section VI) with a summary of
the results and a brief discussion of future applications. In addition
to the appendices that outline the physical motivation for random
Hill's equations, we also present a simple iterative map (Appendix C)
that reproduces our basic results for the growth rates.

\bigskip
\bigskip
\noindent
{\bf II. THE LIMIT OF LARGE FORCING STRENGTH PARAMETER} 
\bigskip 

This section considers the case of large forcing strengths $q_k$ for
the problem with delta function barriers.  This limit applies to
Hill's equations that govern the orbit instabilities in dark matter
halos, embedded young star clusters, and other extended mass
distributions (Appendix A). For the triaxial orbits [4] that
originally motivated this study, for example, the forcing strengths
$q_k \sim 1000$ when the period $\period_k$ and oscillation parameter
$\af_k$ are of order unity.

In this limit, it is useful to factor the matrix ${\mfont M}_k$ that 
connects solutions from cycle to cycle by writing it in the form 
\be 
{\mfont M}_k = h_k {\mfont B}_k \qquad {\rm where} \qquad 
{\mfont B}_k = \left[ \matrix{1 & x_k \phi_k \cr {1/x_k} & 1} \right] \, , 
\label{mbdefine} 
\ee
where $x_k \equiv h_k/g_k$ and where $\phi_k \equiv 1 - 1/h_k^2$. The
ansatz of equation (\ref{mbdefine}) separates the growth rate for this
problem into two parts: $\gamma = \gamma_h + \gamma_B$.  The first
part $\gamma_h$ of the growth rate is given by
\be 
\gamma_h = \lim_{N \to \infty} {1 \over N} \sum_{k=1}^N \log |h_k| \, ,
\label{gammah} 
\ee 
while the remaining part $\gamma_B$ is determined by matrix
multiplication of the matrices ${\mfont B}_k$. In the general problem
[2,3], much of the work thus involves finding the growth rates
$\gamma_B$. In the limit of large $q_k$, however, much simpler --- but
approximate --- forms for the growth rates can be found, as shown below.

\noindent
{\bf Theorem 2.1:} Consider a random Hill's equation with a delta
function barrier. In the limit of large $q_k \gg 1$ and constant
$\af$, the growth rate has the form
\be
\gamma = \left\langle \log \big| 2 h_k \big| \right\rangle \, + 
{\cal O} (1/q_k) \, , 
\label{bigqform} 
\ee 
where the $h_k$ are given by equation (\ref{hk}) and where the angular
brackets denote the expectation value. This form is valid provided
that $\af \ne n^2$, where $n$ is an integer. The width of the zone 
for which equation (\ref{bigqform}) is not valid has order 
$\delta \af = {\cal O} (1/q_k)$. 

\noindent
{\it Proof:} We separate the problem into two pieces according to
equation (\ref{mbdefine}).  The quantities that appear in the matrix
elements in the ${\mfont B}_k$ are the ratios $x_k$ = $h_k / g_k$ and
the correction factors $\phi_k = 1 - 1/h_k^2$.  For the case of delta
function barriers, considered here, the quantities $x_k$ and $\phi_k$
can be written in the form
\be
x_k = { q_k (\pi / \varphi) \sin \varphi - 2 \cos \varphi \over 
q_k (1 + \cos\varphi) + 2 (\varphi/\pi) \sin\varphi } 
\qquad {\rm and} \qquad \phi_k = 1 - 
\left( {2 \varphi \over \pi q_k \sin\varphi 
- 2 \varphi \cos\varphi} \right)^2 \, , 
\label{xk} 
\ee
where we have suppressed the subscripts on the angles $\varphi$ = 
$\sqrt{\af}\pi$.  In the limit of large forcing strength $q_k$, the
ratios $x_k$ become independent of the values of the $q_k$. In
particular, $x_k$ and $\phi_k$ take the asymptotic forms
\be
\lim_{q_k \to \infty} x_k = 
{\pi \sin \varphi \over \varphi (1 + \cos\varphi)} \qquad 
{\rm and} \qquad \lim_{q_k \to \infty} \phi_k = 1 \, . 
\label{bigqlimit}
\ee
For constant $\af$, the angles $\varphi = \sqrt{\af} \pi$ are
also constant, and the $x_k$ are all the same in the limit 
$q_k \to \infty$.  In this same limit, the matrices ${\mfont B}_k$
become constant from cycle to cycle and are denoted here as 
${\mfont B}_0$. These matrices have the simple multiplication property
\be
{\mfont B}_0^2 = 2 {\mfont B}_0 \, , 
\ee
so that the growth rate $\gamma_B = \ln 2$. The remaining part of the
growth rate $\gamma_h$ is given by equation (\ref{gammah}), where the
$h_k$ are given by equation (\ref{hk}). Combining these two results
yields the expression for the growth rate given in equation
(\ref{bigqform}). The correction term is considered below. 

The above derivation of the growth rate $\gamma_B$ is valid as long as
the $x_k$ in equation (\ref{bigqlimit}) remain finite and nonzero.
This requirement leads to the condition that $\af \ne n^2$, where $n$
is an integer. This result also follows from a previous theorem of
Ishii [14,20]. We thus obtain the stated restriction on the range of
validity of the growth rate. To constrain the width of the angular
zone for which this result is not valid, we write $\varphi = n \pi +
\delta\varphi$. Finding the condition for which $|h_k| < 1$, which is
the condition for one cycle to be stable, we find that
\be
\delta \varphi < {4 \sqrt{\af} \over q_k} = {4 n \over q_k} \, . 
\ee
The width of the angular zone thus depends on $q_k$, which varies 
from cycle to cycle, but $\delta \varphi = {\cal O} (1/q_k)$. The 
width of the stability zone for $\af$ is given by 
\be
\delta \af = {2 n \over \pi} \delta \varphi = {8 n^2 \over \pi q_k} \, .
\ee
The width of the zone (in the parameter $\af$) for which the growth 
rate of equation (\ref{bigqform}) fails is thus of order $1/q_k$, 
as claimed. 

In deriving the leading order term in equation (\ref{bigqform}), we
have considered the variables $x_k$ to be constant in the limit of
interest.  In the more general case, the $x_k$ vary from cycle to
cycle (note that varying $\af_k$, not considered here, would also
contribute to variations in $x_k$).  Including these variations leads
to a correction $\Delta \gamma$ to the growth rate. As shown in
Theorem 2 of Ref. [2], this correction can be written in the form
\be
\Delta \gamma = \left \langle \log \big| 1 + x_{k1} / x_{k2} 
\big| \right \rangle - \log 2 \, , 
\label{deltagam} 
\ee
where the $x_{k1}$ and $x_{k2}$ represent two independent samples of
the variable $x_k$ (see equation [\ref{xk}]). Note that $\Delta \gamma
\to 0$ in the limit where the $x_k$ are constant.  In the limit where
the $q_k$ are large, but not infinite, variations in the $x_k$ are
small, and equation (\ref{deltagam}) can be expanded and written in
the approximate form
\be
\Delta \gamma = {1 \over \pi} \left \langle 
{\varphi_{k1} \over q_{k1} \sin \varphi_{k1} } - 
{\varphi_{k2} \over q_{k2} \sin \varphi_{k2} } 
\right \rangle \, + {\cal O} ( q_k^{-2} ) \, . 
\label{xcorrection} 
\ee
The first term is ${\cal O} (1/q_k)$.  If the distributions of $q_k$
and $\varphi_k$ are symmetric, the first term can vanish, and the
correction $\Delta \gamma$ to the growth rate becomes second order in
$1/q_k$.

The result given in equation (\ref{deltagam}) includes the variations
of the $x_k$ but does not take into account possible deviations of the
$\phi_k$ from unity. In order to determine how large these corrections
can be, we consider the case where the correction factors $\phi_k$ are
close to -- but not exactly -- unity.  If $\gamma_B$ is the true growth
rate for multiplication of the matrices ${\mfont B}_k$, and $\gamma_0$
is the growth rate obtained in the limit where $\phi_k \to 1$, then we
denote the difference as $\delta \gamma \equiv \gamma_0 - \gamma_B$. 
For delta function barriers, we can use Theorem 2.3 of Ref. [3] to 
write this correction term in the form 
\be
\delta \gamma = \lim_{N \to \infty} {1 \over N} \sum_{k=1}^N
{ x_k^2 \over (x_k + x_{k+1}) (x_k + x_{k-1}) } 
{4 \varphi_k^2 \over \sin^2 \varphi_k} {1 \over \pi^2 q_k^2} \, , 
\label{deltagrow}
\ee
where $\varphi_k = \sqrt{\af_k} \pi$.  This expression is correct to
leading order in $1/q_k$. As a result, we can use the asymptotic
expressions for the $x_k$ to evaluate $\delta \gamma$ when the $q_k$
are large (but not infinite). In this case, the $x_k$ are independent
of the $q_k$, and the correction $\delta \gamma$ has the order
\be
\delta \gamma = {\cal O} \left( \af_k / q_k^{2} \right) \, , 
\label{dgorder} 
\ee
where $\varphi_k = \sqrt{\af_k} \pi$. As the $q_k$ become large, this
correction decreases, and we recover the growth rates given by
equations (\ref{bigqform}) and (\ref{deltagam}).  If the angle
$\varphi$ is held fixed, then the $x_k$ in equation (\ref{deltagrow})
are identical in the large $q_k$ limit, and this correction 
$\delta \gamma$ to the growth rate reduces to the form
\be
\delta \gamma = \left( {\varphi \over \pi \sin \varphi} \right)^2 
\left\langle {1 \over q_k^2} \right\rangle \, = 
{\af \over \sin^2 \varphi} 
\left\langle {1 \over q_k^2} \right\rangle \, . 
\ee
Note that these results are only valid when $\sin \varphi \ne 0$, 
which in turn requires $\af \ne n^2$ (where $n$ is an integer). 
$\square$  

The above considerations provide corrections for the growth rate of
Theorem 2.1 for cases where the $q_k$ are large, but corrections of
order $1/q_k$ are still relevant. In the opposite limit where 
$q_k \to \infty$, the growth rate can be simplified further:

\noindent
{\bf Corollary 2.1:} For a random Hill's equation with delta
function barriers, in the limit $q_k \to \infty$,
the growth rate approaches the form 
\be
\gamma = \left\langle \log \bigg| {q_k \over \sqrt{\af}}
\sin \sqrt{\af} \pi \bigg| \right\rangle \, . 
\label{infqform} 
\ee 
This form is valid for $\af \ne n^2$, where $n$ is an integer.

\noindent
{\it Proof:} This limit represents a stronger version of the
conditions for which equation (\ref{bigqform}) is valid. Starting with
equation (\ref{hk}), the matrix elements $h_k$ approach the following
form in the limit $q_k \to \infty$:
\be
h_k \to - {q_k \over 2 \sqrt{\af}} \sin \varphi \, . 
\label{hlimit} 
\ee
Using this expression in equation (\ref{bigqform}), we obtain the
claimed form for the growth rate given in equation (\ref{infqform}).
$\square$

Notice that the difference between the asymptotic form of the growth
rate from equation (\ref{infqform}) and the approximation of equation
(\ref{bigqform}) is first order in $1/q_k$. For comparison, the
corrections due to variations in the $x_k$ are first order for
asymmetric variations and second order for the symmetric case (see
equation [\ref{xcorrection}]).  The corrections due to the departure
of the $\phi_k$ from unity are second order (equations
[\ref{deltagrow}] and [\ref{dgorder}]).

\begin{figure} 
\centering
\includegraphics[scale=0.75]{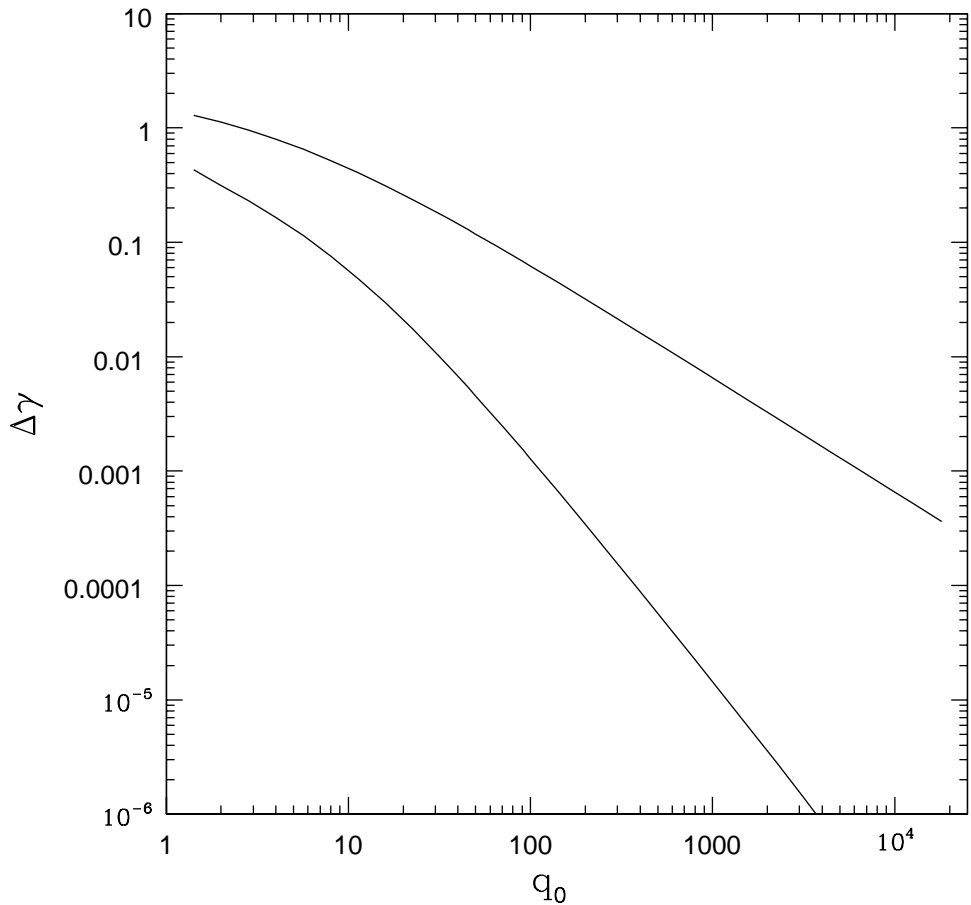} 
\caption{Degree of validity for the approximations to the growth rates
for Hill's equation in the delta function limit with large forcing
parameters. The lower curve shows the difference between full growth
rate calculated from matrix multiplication and that calculated from
Theorem 2.1 (equation [\ref{bigqform}]). The upper curve shows the
difference between the full growth rate and that calculated from the
more extreme approximation of Corollary 2.1 (equation 
[\ref{infqform}]).  Here, the values of $\af$ are fixed and the values
of $q_k$ fluctuate according to $q_k$ = $(1 + \xi_k) q_0$, where
$\xi_k$ is a uniform random variable and the constant $q_0$ provides 
a measure of the fluctuation amplitude (where $\langle q_k \rangle$ 
= $3 q_0 / 2$). }
\label{fig:qbigerror} 
\end{figure}

Figure \ref{fig:qbigerror} illustrates the validity of the
approximations derived in Theorem 2.1 and its corollary.  Here we take
Hill's equation in the delta function limit, with fixed oscillation
parameter $\af$. We then let the forcing strength vary according to
$q_k$ = $(1 + \xi_k) q_0$, where $\xi_k$ is a uniformly distributed
random variable and the constant $q_0$ determines the amplitude of the
fluctuations. The growth rates are calculated using three successive
approximations: (a) the full matrix multiplication scheme from
equation (\ref{mapzero}) with growth rate given by equation
(\ref{growbasic}), (b) the approximation of equation (\ref{bigqform})
which is valid in the limit of large forcing strength $q_k$, and (c)
the more extreme approximation of equation (\ref{infqform}), which is
valid in the limit $q_k \to \infty$.  The upper curve in the figure
shows the difference between the full growth rate of equation
(\ref{growbasic}) and that of equation (\ref{infqform}); the lower
curve shows the difference between the full growth rate and that of
equation (\ref{bigqform}).  Both approximations work well for large
$q_k$, measured here using $q_0$, where ``large'' means $q_0$ greater
than $\sim100$. Both curves approach power-law forms, with
well-defined slopes, showing that the approximation of Theorem 2.1 is
valid to second order in $1/q_0$, whereas the more extreme
approximation of Corollary 2.1 is only accurate to first order in
$1/q_0$.

\bigskip
\bigskip
\noindent
{\bf III. THE LIMIT OF SMALL FORCING STRENGTH PARAMETER} 
\bigskip

This section considers the limiting regime where the forcing strength
$q_k \ll 1$. This limit is expected to be applicable to the reheating
problem after an inflationary epoch (Appendix B). The reheating phase
takes place over many oscillations of the inflaton field and hence
many cycles of the corresponding Hill's equation.  As a result, the
fluctuations (given by the magnitude of the $q_k$) must be relatively
small.

In general, when the forcing parameter $q_k$ is small, solutions to
Hill's equation tend to be stable in the classical regime, i.e., where
the parameters do not vary from cycle to cycle.  However, variations
in the parameters $(\af_k, q_k)$ allow for unstable solutions, even if
the growth rate would vanish in the absence of fluctuations. For the
case of delta function barriers, the parameter $q_k / \sqrt{\af_k}$
must be small for stability, so that classically stable solutions can
also arise in the limit of large oscillation frequency $\af_k$ (see
equation [\ref{deltaprin}]).  Here we find the growth rate for a
random Hill's equation in the limit of small forcing strength for the
case of delta function barriers:

\noindent
{\bf Theorem 3.1:} Consider a random Hill's equation (\ref{basic})
with a delta function barrier so that $\qhat(t)$ is given by equation
(\ref{qhatdelta}).  In the limit of small $q_k \ll 1$, fixed $\af$,
and symmetric variations in the $q_k$, the growth rate approaches 
the form:
\be
\gamma = \log \left[ 1 + \langle q_k^2 \rangle / 8 \af \right] \, , 
\label{quadform} 
\ee 
where the angular brackets denote expectation values. This form is
valid for all $\af > 0$ except for narrow bands of stability centered
on square integer values $\af = n^2$ ($n \in {\mfont Z}$). The growth
rate vanishes at these values of $\af$ and the width $\delta \af$ of
the bands is given by
\be
\delta \af \approx 2 q_k / \pi \, . 
\label{zonewidth} 
\ee

\noindent
{\it Proof:} In the limit of delta function barriers, the principal
solutions are given by equation (\ref{deltaprin}). For the case of
small forcing parameters $q_k \ll 1$, the $h_k$ are less than unity
except for the narrow zones of parameter space defined by the
condition $\af$ = $n^2$ and by equation (\ref{zonewidth}). As a
result, we can rewrite the matrix elements $h_k$ in the form
\be
h_k \equiv \cos \theta_k \, . 
\label{deftheta} 
\ee
The transformation matrix of equation (\ref{mapzero}) 
can be written in the form 
\be
{\mfont M}_k = \left[ 
\matrix{\cos \theta_k & - L_k \sin \theta_k \cr 
\sin \theta_k / L_k & \cos \theta_k } \right] \, , 
\label{mapellipse} 
\ee
where the parameter $L_k$ is defined by 
\be
L_k \equiv {\sin \theta_k \over g_k} \, = \, - 
{\sin\varphi \left[1 + (q_k/\sqrt{\af}) \cot \varphi -
(q_k^2/4 \af) \sin^2 \varphi \right]^{1/2} \over 
\sqrt{\af} \sin\varphi + (q_k/2) (1 + \cos\varphi) } \, \, . 
\label{lkdef} 
\ee
To leading order in $q_k \ll 1$, $L_k = L_0 = -1/\sqrt{\af}$, where
$L_0$ is a constant. If we write $L_k = L_0 (1 + \eta_k)$, the
perturbations can be written in the form
\be
\eta_k = {\sin \theta_k \over 
\sin\varphi + (q_k/2 \sqrt{\af}) (1 + \cos\varphi) } - 1 = 
- {q_k \over 2 \sqrt{\af} \sin\varphi} \, + {\cal O} (q_k^2) \, , 
\label{etadef} 
\ee
where the second equality defines the leading order expression. 

As shown below, the product $(\eta_k \sin\theta_k)$ appears in the
expression for the growth rate and is thus the quantity of interest.
Since $\eta_k$ is first order in $q_k$, we can use the leading order
expression for $\sin \theta_k$ to evaluate the product.  For small
$q_k \ll 1$ trigonometric identities imply the following
transformation between the angle $\varphi_k$ and the angle $\theta_k$:
\be
\theta_k = \varphi + {q_k \over 2 \sqrt{\af} } \, .
\label{sumangle} 
\ee 
As a result, $\sin\theta_k = \sin\varphi + {\cal O}(q_k)$, so that 
$(\eta_k \sin\theta_k)$ = $-q_k / (2 \sqrt{\af})$ to leading order.

Next we expand the transformation matrix of equation (\ref{mapellipse}) 
into two parts, 
\be
{\mfont M}_k = {\mfont M}_{0k} (\theta_k; L_0) + {\mfont M}_{1k} \, , 
\label{matsum} 
\ee
where the first term is an elliptical rotation matrix with constant
length parameter $L_0$ and where the second term has the form
\be 
{\mfont M}_{1k} = \, - \eta_k \,  \sin \theta_k \, 
\left[ \matrix{0 & L_0 \cr 
1 / [L_0 (1 + \eta_k)] & 0 } \right] \, .
\ee
Note that the first term in equation (\ref{matsum}) is stable under
matrix multiplication. Notice also that the second term ${\mfont
M}_{1k}$, as written, includes the full correction (with no
approximations).

As shown in the following analysis, the first non-vanishing contribution 
to the growth rate is second order in $\eta_k$. As a result, we expand 
the product of $N$ matrices ${\mfont M}_k$,  
\be
{\mfont M}_k^{(N)} = \left( {\mfont M}_{0k} + 
{\mfont M}_{1k} \right)^N \, , 
\ee
including all terms to second order in the matrix ${\mfont M}_{1k}$, 
\be
{\mfont M}_k^{(N)} = {\mfont M}_{0k}^{(N)} + 
\sum_{k=1}^N {\mfont P}_k^N + 
\sum_{k,\ell} {\mfont Q}_{k\ell}^N \, \, . 
\label{expansion} 
\ee
The first sum includes partial product matrices of the form  
\be 
{\mfont P}_k^N = \left\{ \prod_{j=k+1}^N {\mfont M}_{j0} \right\} \, 
{\mfont M}_{1k}  \left\{ \prod_{j=1}^{k-1} {\mfont M}_{j0} \right\} \, 
= \, {\mfont E}_0 (a_k; L_0) \, 
{\mfont M}_{1k} \, {\mfont E}_0 (b_k; L_0) \, . 
\label{pfull} 
\ee
In the second equality we have evaluated the products using the
properties of the elliptical rotation matrices, denoted here as 
${\mfont E}_0$, and we have defined the composite angles 
\be
a_k \equiv \sum_{j=k+1}^N \theta_j
\qquad {\rm and} \qquad 
b_k \equiv \sum_{j=1}^{(k-1)} \theta_j \, . 
\label{abdef} 
\ee
The second sum in the expansion of equation (\ref{expansion}) involves
partial product matrices with the form
\be 
{\mfont Q}_{k\ell}^N = \left\{ \prod_{j=k+1}^N {\mfont M}_{0j} \right\} \, 
{\mfont M}_{1k} \left\{ \prod_{j=\ell + 1}^{k-1} {\mfont M}_{0j} \right\} \, 
{\mfont M}_{\ell 1} \left\{ \prod_{j=1}^{\ell-1} {\mfont M}_{0j} \right\} \, .
\ee
This second sum includes all possible products of the above form,
i.e., all possible locations of the two matrices that are of type
${\mfont M}_{1k}$ rather than ${\mfont M}_{0k}$.  By construction,
each matrix ${\mfont Q}_{k\ell}$ contains two factors of the random
variable so that ${\mfont Q}_{k\ell} \propto \eta_k \eta_\ell$, where
the $\eta_k$ and $\eta_\ell$ are independent realizations and hence
are uncorrelated. The matrix elements from the third term in equation
(\ref{expansion}), the sum over the ${\mfont Q}_{k\ell}^N$, must thus
vanish in the limit $N \to \infty$ (see also [3]). As a result, we
only need to consider the contribution from the first sum in equation
(\ref{expansion}). In this sum, the first order terms, those
proportional to $\eta_k$, will also vanish in the limit
$N \to \infty$.  We thus need to include the second order terms in the
first sum.  Using the result of equation (\ref{pfull}) and expanding
to second order in $\eta_j$, we thus obtain 
\be
\sum_{k=1}^N {\mfont P}_k^N = \sum_{k=1}^N \eta_k^2 \sin\theta_k 
\left[ \matrix{ - \sin a_k \cos b_k & L_0 \sin a_k \sin b_k \cr 
(1/L_0) \cos a_k \cos b_k & - \cos a_k \sin b_k } \right] \, . 
\ee 
After using this result in the expansion of equation (\ref{expansion}), 
and evaluating the product ${\mfont M}_{0k}^N$, the eigenvalue
$\eigenvalue$ for the full product matrix after $N$ steps is given by
\be
\eigenvalue^2 - 2 \eigenvalue \cos \theta_N + 1 + 
\sum_{k=1}^N \left\{ \eta_k^2 \sin\theta_k 
\left[ \left( \eigenvalue - \cos \theta_N \right) \sin \alpha_k + 
\sin \theta_N \cos \alpha_k \right] \right\}  = 0 \, , 
\label{eigensimple} 
\ee    
where we have defined $\alpha_k \equiv a_k + b_k$ 
(see equation [\ref{abdef}]), and where 
\be
\theta_N \equiv \sum_{j=1}^N \theta_k \, . 
\ee
The zeroth order contribution to the eigenvalue is given by
\be
\eigenvalue_0 = \cos \theta_N \pm i \sin \theta_N \,  ,
\label{eigenzero} 
\ee
and the leading order correction is given by 
\be
\eigenvalue_2 = {\pm i \over 2} \sum_{k=1}^N 
\left\{  \eta_k^2 \sin \theta_k \left( 
\cos\alpha_k \pm i \sin \alpha_k \right) \right\}  \, . 
\ee
The magnitude of the full eigenvalue, $\eigenvalue = \eigenvalue_0  + \eigenvalue_2$, 
is then given by 
\be 
|\eigenvalue| = 1 + {1 \over 2} \sum_{k=1}^N \eta_k^2 \sin^2 \theta_k \, 
= 1 + {1 \over 2} N \langle \eta_k^2 \sin^2 \theta_k \rangle \, , 
\ee 
where the second equality is valid in the limit $N \to \infty$.  To 
leading order in $\eta_k$, this expression can be rewritten in the form
\be 
|\eigenvalue| = \left[ 1 + {1 \over 2} 
\langle \eta_k^2 \sin^2 \theta_k \rangle \right]^N \, . 
\ee 
The corresponding growth rate thus becomes 
\be
\gamma = \lim_{N \to \infty} {1 \over N} \log |\eigenvalue| = 
\log \left[ 1 + {1 \over 2} 
\langle \eta_k^2 \sin^2 \theta_k \rangle \right] \, . 
\ee 
Using equation (\ref{etadef}) to determine $\eta_k$, we obtain 
the expression claimed in equation (\ref{quadform}). 

To prove the second part of this theorem, we note that the expansion of
equation (\ref{etadef}) is no longer valid when the second term in the
denominator of equation (\ref{lkdef}) dominates the first. The
condition for the expansion to fail can then be written in the form
\be
\left| {q_k \over \sqrt{\af}} {\cos^2 (\varphi/2) \over \sin\varphi} 
\right| \sim 1 \, . 
\ee
The left hand side of this equation blows up when $\sin\varphi=0$,
which occurs when $\varphi = n \pi$ and $n$ is an integer;
equivalently, this singularity occurs when $\sqrt{\af}$ is an integer
(and $\af$ is a square integer).  Near these square integer values of 
$\af$ of interest, we can write
\be
\varphi = \sqrt{\af} \pi \equiv n\pi + \delta \varphi \, , 
\ee
where the second equality defines $\delta \varphi$. Combining the
above two results implies that
\be
{\pi q_k /2 \over n \pi + \delta\varphi} 
\left[ 1 + (-1)^n \cos (\delta\varphi) \right] 
\sim \delta\varphi \, . 
\ee
For $n$ even, we thus obtain $\delta \varphi \sim q_k/n$ to leading
order.  By definition, $\delta \af = 2 n (\delta\varphi)/\pi$, so
the width of the interval where equation (\ref{quadform}) fails is
given by $\delta \af \sim 2 q_k / \pi$, in agreement with equation
(\ref{zonewidth}). This derivation applies to even integers $n$.  For
the case of odd $n$, one can derive the analogous result.  $\square$

\begin{figure} 
\centering
\includegraphics[scale=0.75]{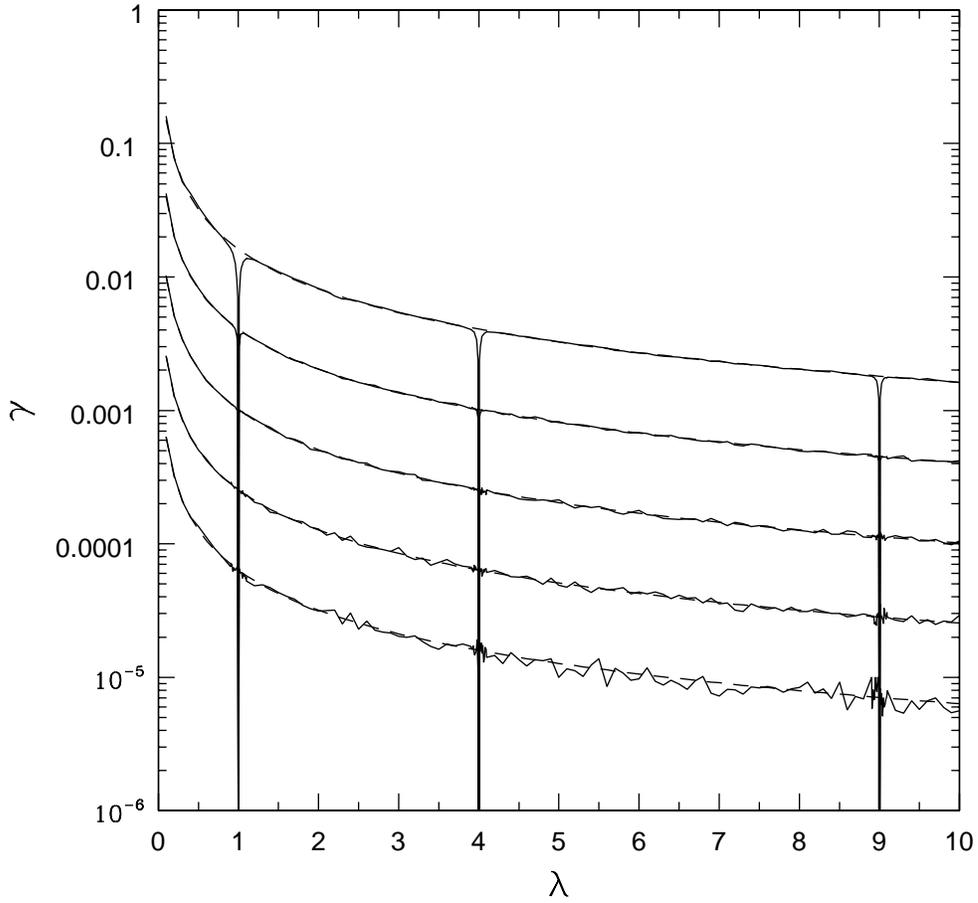} 
\caption{Growth rates for Hill's equation in the delta function
limit for fixed values of $\af$ and fluctuating values of $q_k$.
The five curves shown here correspond to five values of the fluctuation 
amplitude $q_0$, where the $q_k = q_0 \, \xi_k$, where $\xi_k$ is a 
uniformly distributed random variable $-1 \le \xi_k \le 1$. For the 
five curves shown, the amplitudes are given by $q_0$ = 10/2$^\ell$ 
for $\ell$ = 4, 5, 6, 7, 8. The dashed lines show the limiting form 
for the growth rate from equation (\ref{quadform}). } 
\label{fig:deltaqk} 
\end{figure}

\begin{figure} 
\centering
\includegraphics[scale=0.75]{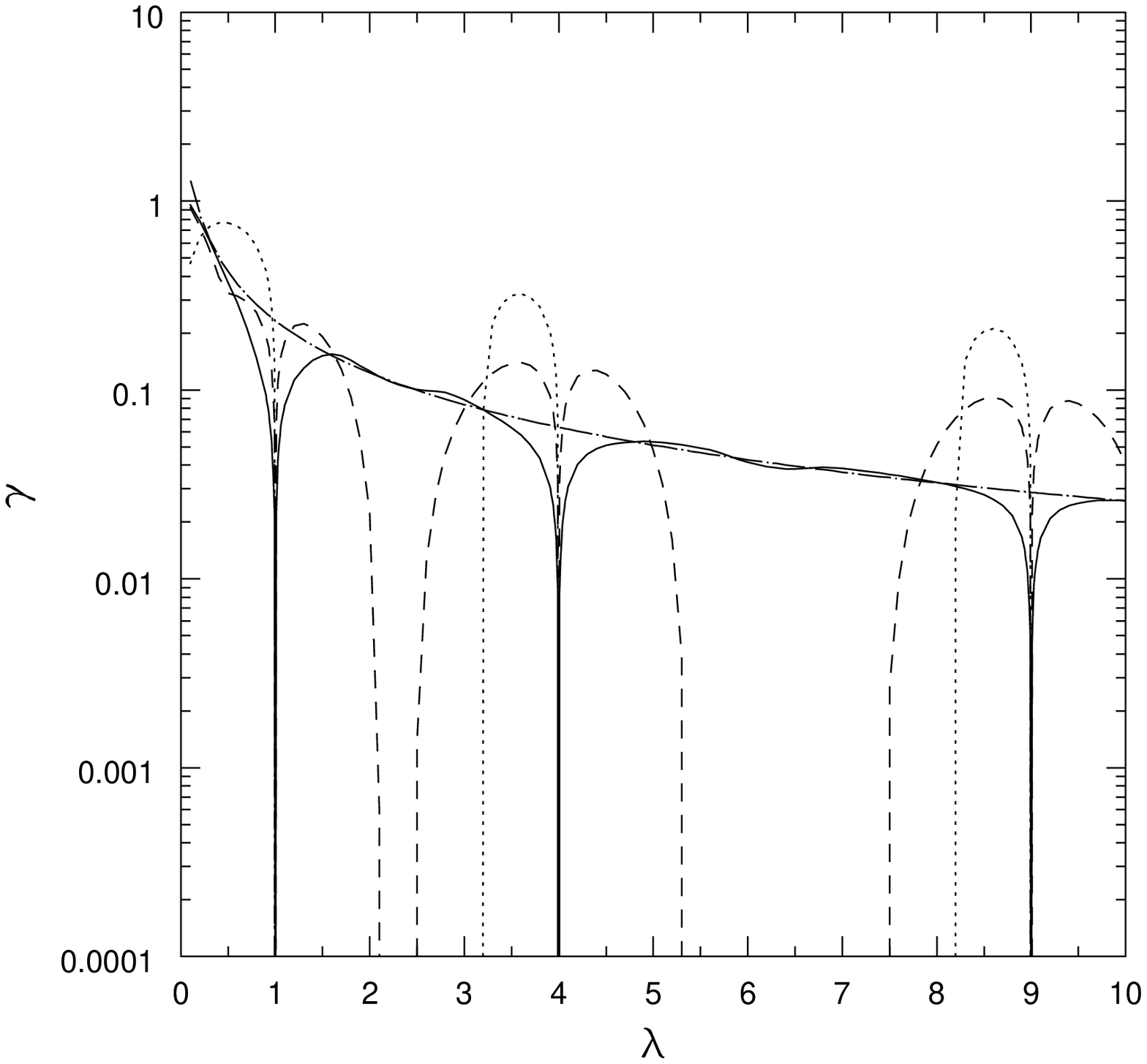} 
\caption{Comparison of growth rates for Hill's equation in the
delta function limit for fixed $\af$ and fluctuating values 
of the forcing strength $q_k$.  The solid curve corresponds to 
the exact result from matrix multiplication using a fluctuation
amplitude $q_0$ = 2.5.  The dot-dashed curve shows the approximation
developed in this section (equation [\ref{quadform}]). The dashed
curve shows the growth rate $\gamma_\infty$ that results from an
average of the growth rates for individual cycles (the asymptotic 
growth rate). Finally, the dotted curve shows the growth rate
resulting from a constant value of the forcing strength $q$ =
$q_0/2$. }
\label{fig:compare} 
\end{figure}

To illustrate this set of results, we present the following numerical
experiment: The natural oscillation frequency, as set by the parameter
$\af_k$, is fixed at a constant value.  The forcing strength is then
allowed to vary according to the ansatz
\be
q_k = q_0 \xi_k \, , 
\label{qvary} 
\ee 
where $q_0$ is a fixed amplitude and $\xi_k$ is a uniformly
distributed random variable with $-1 \le \xi_k \le 1$. In the absence
of the fluctuations, Hill's equation would have bands of stability and
bands of instability in the ($\af$-$q$) plane of parameters [1,2,19].  
However, with the cycle to cycle variations of the forcing strength
given by equation (\ref{qvary}), the bands of stability essentially
disappear.  Figure \ref{fig:deltaqk} shows the growth rate plotted as
a function of $\af$ for a collection of amplitudes $q_0$ (where the
amplitudes are equally spaced logarithmically, so that $q_0$ =
$10/2^\ell$ for $\ell$ = 4,5,6,7,8).  Even for extremely small values
of $q_0$ (and hence correspondingly small $q_k$), the growth rates are
nonzero. The growth rate does vanish for particular values of the
frequency parameter $\af$, where $\af$ = $n^2$ and $n$ is an integer. 
At these particular frequencies, $\sin\theta$ = 0 and $h_k =
\cos\theta = \pm 1$ = {\sl constant}. Notice also that in the limit
$q_0 \ll 1$, this incarnation of the random Hill's equation is
equivalent to a simple harmonic oscillator with frequency $\af$ and
perturbative noise; this result -- that noise leads to instability or
can speed up instability -- has analogs in previous work [9].

For the particular choice of $q_k$ used in Figure \ref{fig:deltaqk},
$\langle q_k^2 \rangle = q_0^2/3$. The dashed curves in Figure
\ref{fig:deltaqk} show the approximation of equation (\ref{quadform})
for the growth rate. In this case, the approximate form actually gives
more accurate results than direct matrix multiplication (solid curves)
due to incomplete sampling in the latter. However, the expression of
equation (\ref{quadform}) does not account for the vanishing of the
growth rate for particular values of $\af$.

As another illustration of how random variables change the landscape
of parameter space, Figure \ref{fig:compare} shows the growth rates as
a function of (fixed) $\af$ for several cases. The solid curve shows
the result from full matrix multiplication, with the same sampling of
$q_k$ as used in Figure \ref{fig:deltaqk}, with value $q_0$ = 2.5.
The dot-dashed curve shows the prediction from the approximation of
equation (\ref{quadform}). The approximation works well except near
the integer square values of the frequency $\af$ where the growth rate
vanishes.  However, since the $q_k$ are of order unity, rather than
fully in the regime $q_k \ll 1$, the assumptions for the validity of
equation (\ref{quadform}) are not completely satisfied. As a result,
the small amplitude oscillations of the approximate result (dot-dashed
curve) about the true growth rate (solid curve) are real. In Ref. [2]
we defined the asymptotic growth rate $\gamma_\infty$ to be the growth
rate for a random Hill's equation resulting from an appropriate
average of the growth rates for the individual cycles; this quantity
is plotted as the dashed curve in Figure \ref{fig:compare}. For this
set of $q_k$ values, the amplitude is $q_0$, so the mean of the
forcing parameter magnitude is $q_0/2$; for comparison, the dotted
curve shows the growth rates for the classical problem (no random
variables) with $q = q_0/2$.  The dotted curves thus delineate the
regions of stability and instability that characterize the parameter
space of Hill's equation.  Note that the asymptotic growth rate
$\gamma_\infty$ is sometimes larger and sometimes smaller than the
true growth rate. As a general rule, one finds $\gamma_\infty >
\gamma$ in or near the portions of parameter space for which the
classical problem (fixed $q$) is unstable; for regimes in which the
systems is classically stable, however, the opposite holds so that
$\gamma > \gamma_\infty$.

\bigskip
\bigskip
\noindent
{\bf IV. FOKKER-PLANCK APPROACH} 
\bigskip

For the case of constant oscillation frequency parameter $\af$ and
sufficiently small forcing strengths $q_k$, Hill's equation in the
limit of delta function barriers can be described by a Fokker-Planck
equation of conventional form. Following standard methods, we define
the velocity $V$ and diffusion constant $D$ according to
\be 
V \equiv {d y \over dt} \qquad {\rm and} \qquad 
D \equiv { \langle q_k^2 \rangle \over \pi} \, , 
\ee
where $\pi$ is the period of the forcing intervals. The Fokker-Planck 
equation for the evolution of the distribution $P(y,V,t)$ of phase space 
variables thus becomes 
\be
{\partial P \over \partial t} + V {\partial P \over \partial y}
- \af y {\partial P \over \partial V} = {D \over 2} y^2 
{\partial^2 P \over \partial V^2} \, .
\label{fokplanck} 
\ee
Notice that including variations in the $\af_k$, or working in the
regime of large forcing parameters $q_k$ (e.g., the highly unstable
limit), would require additional terms in equation (\ref{fokplanck}).

In order to reduce the complexity of equation (\ref{fokplanck}), it
would be useful to average over one of the independent variables. In
this case, however, both $V$ and $y$ are on a nearly equal footing. In
this problem, the system acts like a simple harmonic oscillator,
except at the delta function barriers where its energy jumps due to
the forcing.  We thus transform into a type of ``polar coordinates''
[20] in which the energy plays the role of the radial coordinate; 
specifically we define
\be 
E \equiv {1 \over 2} \left( V^2 + \af y^2 \right) \qquad {\rm and} 
\qquad \eangle = \tan^{-1} \left( \sqrt{\af} y/V \right) \, , 
\ee
where the corresponding inverse transformation takes the form 
\be
V = 2 \sqrt{E} \cos \eangle \qquad {\rm and} \qquad 
\sqrt{\af} y = 2 \sqrt{E} \sin \eangle \, . 
\ee
In terms of the new variables $(E, \eangle)$, 
the Fokker-Planck equation becomes
$$
{\partial P \over \partial t} + 
\sqrt{\af} {\partial P \over \partial \eangle} = 
{2 D \over \af} E \sin^2 \eangle \Biggl\{ 
{\partial P \over \partial E} + 
4 E \cos^2 \eangle {\partial P^2 \over \partial E^2} \qquad 
\qquad \qquad \qquad \qquad 
$$
\be
\qquad \qquad \qquad \qquad 
+ {\sin 2\eangle \over E} {\partial P \over \partial \eangle} + 
{\sin^2 \eangle \over E} {\partial P^2 \over \partial \eangle^2} -
2 \sin 2\eangle {\partial^2 P \over \partial E \partial \eangle} 
\Biggr\} \, . 
\label{fpfull} 
\ee
If we now average over the angular variable $\eangle$, the 
partial derivative terms with respect to $\eangle$ vanish, 
and the Fokker-Planck equation simplifies to the form 
\be
{\partial P \over \partial t} = 
{2 D \over \af} E \left\{ \langle \sin^2 \eangle \rangle 
{\partial P \over \partial E} + 4 E 
\langle \sin^2 \eangle \cos^2 \eangle \rangle 
{\partial P^2 \over \partial E^2} \right\} = 
{D \over \af} \left\{ E {\partial P \over \partial E} 
+ E^2 {\partial P^2 \over \partial E^2} \right\} \, . 
\label{fpenergy} 
\ee
Next we change variables again, by defining 
\be
\mu \equiv \log E \, . 
\ee
Note that we must take the logarithm of a dimensionless quantity.
Given the form of equation (\ref{fpenergy}), however, it is
straightforward to introduce a dimensionless energy ${\widetilde E}
\equiv E/E_0$ before changing to the logarithmic form. The resulting 
Fokker-Planck equation thus becomes an ordinary diffusion equation
\be
{\partial P \over \partial t} = 
{D \over \af} {\partial P^2 \over \partial \mu^2} \, . 
\ee
This diffusion equation has the normalized solution 
\be
P(\mu,t) = \left( {\af \over 4 \pi D t} \right)^{1/2} 
\exp \left[ - {\mu^2 \af \over 4 D t} \right] \, , 
\ee
which is appropriate for the boundary condition $P(\mu, t=0) =
\delta(\mu)$, i.e., the system starts out with $\mu$ = 0 or energy 
$E = E_0$.  The solutions $y(t)$ are oscillatory, but growing (in 
general).  In order to extract a growth rate from this Fokker-Planck
treatment of the problem, we first determine the expectation value of
$y^2$, which takes the form
\be
\langle y^2 \rangle = \int_{-\infty}^{\infty} P (\mu,t) d\mu 
{4 E \sin^2 \eangle \over \af} = {2 \over \af} 
\exp \left[ D t / \af \right] \, . 
\ee
To obtain the final expression, have averaged over the angle 
$\eangle$, so that $2 \sin^2 \eangle = 1$. Next we assume that 
the amplitude $|y|$ of the solution can be characterized by 
the relationship 
\be
|y| \approx \langle y^2 \rangle^{1/2} \propto 
\exp \left[ \gamma_{\rm fp} t \right] \, , 
\ee
where $\gamma_{\rm fp}$ is the growth rate resulting from this
Fokker-Planck approach. The resulting estimate for the growth rate is
\be
\gamma_{\rm fp} = {D \over 2 \af} = 
{\langle q_k^2 \rangle \over 2 \pi \af} \, . 
\label{fpgrowth} 
\ee
This growth rate is similar, but not identical to, that given by
Theorem 3.1 for the case of constant frequency parameter $\af$, delta
function barriers, and in the limit of small forcing parameters $q_k$.
The functional dependence $\gamma \propto q_k^2/\af$ is the same, only
the numerical coefficient differs. In order to derive the result
(\ref{fpgrowth}), however, we have averaged the Fokker-Planck equation
(\ref{fpfull}), and this procedure can produce such a numerical
difference.

This treatment using the Fokker-Planck equation thus provides a good
description of the problem for the case of small $q_k$ and constant
$\af$. For varying values of $\af_k$, additional diffusive terms must
be included. For the case of large $q_k$, however, the Fokker-Planck
approach does not naturally reproduce the results obtained here using
direct methods. The derivation of equation (\ref{fokplanck}) involves
truncating a series of terms in powers of $q_k^n$ [6], and such a
truncation is only valid for sufficiently small forcing strengths. In
the highly unstable limit (large $q_k$), one must either use a highly
modified form of the Fokker-Planck equation or abandon this approach
altogether.

\bigskip
\bigskip
\noindent
{\bf V. TRANSFORMATION FROM HILL'S EQUATION PARAMETERS TO MATRIX ELEMENTS} 
\bigskip

For Hill's equations with delta function barriers, we can write the
matrix elements $h_k$ and $g_k$ in the form given by equation
(\ref{deltaprin}) above. With these results in hand, we can directly
construct the relationship between the fundamental parameters $(\af_k,
q_k)$ appearing in the original differential equation (\ref{basic})
and the moments of the distributions of the matrix elements.

We start by considering the case where the angle $\varphi_k$ is held
fixed, but the forcing strength $q_k$ is allowed to vary. For ease of
notation, we suppress the subscripts on the angle $\varphi$ and the
forcing strength $q$.  The mean (first moment) of the matrix element
$h_k$ is then given by 
\be
\langle h_k \rangle = \cos \varphi - {\pi \over 2} \langle q \rangle 
{\sin\varphi \over \varphi} \, . 
\label{hmomentoneq} 
\ee
Similarly, the second moment takes the form 
\be
\langle h_k^2 \rangle = \cos^2 \varphi + {\pi^2 \over 4} 
\langle q^2 \rangle \left( {\sin\varphi \over \varphi} \right)^2 - 
\pi{\cos\varphi \sin\varphi \over \varphi} \langle q \rangle \, . 
\label{hmomenttwoq} 
\ee
The variance is thus given by the expression 
\be
\sigma_h^2 = {\pi^2 \over 4} \left( {\sin\varphi \over \varphi} \right)^2
\left\{ \langle q^2 \rangle - \langle q \rangle^2 \right\} \, = 
{\pi^2 \over 4} \left( {\sin\varphi \over \varphi} \right)^2 \sigma_q^2 \, , 
\ee
where the second equality defines the variance of the distribution of
the forcing strength $q$.  As a result, this limiting case provides a 
simple relationship between the width of the distribution of forcing
strength $q$ and the width of the distribution of the resulting matrix
element $h_k$, i.e., 
\be
\sigma_h = {\pi \, \sin\varphi \over 2\varphi} \, \sigma_q \, . 
\label{sigmah} 
\ee

Next we consider the opposite case in which the forcing strength $q$ is 
held fixed and the angle $\varphi$ varies over a range. Here we consider a 
range of angles so that expectation values are taken via the operator 
\be
\langle \dots \rangle \equiv {1 \over \aint} 
\int_0^\aint d\varphi \dots \, , 
\label{angleop} 
\ee 
which holds for any given quantity in the brackets. Notice that we are
using a distribution that is uniform in the variable $\varphi$. Since
$\varphi \propto \sqrt{\af}$, this distribution is not uniform in the
variable $\af$, although an analogous analysis could be done for that
case. For this choice of distribution, the first moment of the matrix
element is then given by the expression
\be
\langle h_k \rangle = {\sin \aint \over \aint} - 
{\pi q \over 2 \aint} \sint (\aint) \, ,
\label{hmomentone} 
\ee
where $\sint(\aint)$ is the sine integral [1]. The second moment 
is given by 
\be
\langle h_k^2 \rangle = {1 \over 2} + {\sin2\aint \over 4\aint} 
+ \left( {\pi q \over 2} \right)^2 \left\{ {1 \over \aint} \sint(2\aint) - 
{\sin^2 \aint \over \aint^2} \right\} - {\pi q \over 2 \aint} \sint(2\aint) \, . 
\label{hmomenttwo} 
\ee
These (general) expressions can be simplified by choosing the angle
interval to be have the form $\aint = 2 \pi m$, where $m$ is an integer. 
In this case, the corresponding variance reduces to the form 
\be
\sigma_h^2 = {1 \over 2} + {\pi q \over 2 \aint} 
\left( {\pi q \over 2} - 1 \right) \sint(2\aint) - 
\left( {\pi q \over 2\aint} \right)^2 \sint^2 (\aint) \, .
\ee
Next we note that these results simplify further in the limit $\aint
\to \infty$, i.e., when we allow the angle $\varphi$ to vary uniformly
over the entire positive real line. In this limit we find
\be
\lim_{\aint \to \infty} \langle h_k \rangle = 0 \, , \qquad
\lim_{\aint \to \infty} \langle h_k^2 \rangle = {1 \over 2} \, , \qquad 
{\rm and} \qquad \lim_{\aint \to \infty} \sigma_h = {\sqrt{2} \over 2} \, . 
\ee

As long as the distributions of the angle $\varphi$ and that of the
forcing strength parameter $q$ are independent, the expressions
derived above for the moments can be generalized to include both
distributions in a straightforward manner.  In particular, we continue
to use the distribution of equation (\ref{angleop}) for the angle, and
the same (unspecified) distribution for the forcing strength as above.
In this case, we must integrate over both the forcing strength $q$ and
the angle $\varphi$. By performing the integrals over $q$ first, we
obtain expressions analogous to equations (\ref{hmomentoneq}) and
(\ref{hmomenttwoq}) for the first two moments. With these results in
hand, we then average over the distribution of angle using equation
(\ref{angleop}).  This procedure produces expressions of the forms
given by equations (\ref{hmomentone}) and (\ref{hmomenttwo}), with $q$
replaced by $\langle q \rangle$ and with $q^2$ replaced by $\langle
q^2 \rangle$. For the case in which $\aint$ = $2 \pi m$, the resulting
expressions for the first two moments take the form
\be
\langle h_k \rangle = - {\pi \over 2 \aint} \sint (\aint) 
\langle q \rangle \qquad {\rm and} \qquad 
\langle h_k^2 \rangle = {1 \over 2} + 
{\pi^2 \over 4 \aint} \sint(2\aint) \langle q^2 \rangle 
- {\pi \over 2 \aint} \sint(2\aint) \langle q \rangle \, . 
\ee
The corresponding variance is thus given by 
\be
\sigma_h^2 = {1 \over 2} + {\pi^2 \over 4 \aint} \sint(2\aint) 
\left\{ \langle q^2 \rangle - {2 \over \pi} \langle q \rangle \right\} 
- {\pi^2 \over 4 \aint^2} \langle q \rangle^2 \sint^2 (\aint) \, . 
\ee
If we consider the limit where both $\langle q \rangle$ and $\aint$ 
are large, the variance of $h_k$ has the order 
\be
\sigma_h^2 = {1 \over 2} + {\cal O} 
\left( {\sigma_q^2 / \aint} \right) \, . 
\label{orderhsig} 
\ee

Next we consider the analogous calculation for the matrix element
$g_k$.  For the limiting case in which the angle is held fixed and the
forcing strength varies, the first two moments of the distribution are
given by
\be
\langle g_k \rangle = - {1 \over \pi} \varphi \sin \varphi - 
{1 \over 2} \left( 1 + \cos\varphi \right) \langle q \rangle \, , 
\ee
and 
\be
\langle g_k^2 \rangle = {1 \over \pi^2} \varphi^2 \sin^2 \varphi + 
{1 \over 4} \left( 1 + \cos\varphi \right)^2 \langle q^2 \rangle - 
{1 \over \pi} \varphi \sin \varphi \left( 1 + \cos\varphi \right) 
\langle q \rangle \, .
\ee
The corresponding variance thus takes the form 
\be
\sigma_g^2 = {1 \over 4} \left( 1 + \cos\varphi \right)^2 
\left\{ \langle q^2 \rangle - \langle q \rangle^2 \right\} \, . 
\ee
This result can be rewritten in a manner analogous to that found 
for the other principal solution (equation [\ref{sigmah}]), i.e.,
\be
\sigma_g = {1 \over 2} \left( 1 + \cos\varphi \right) \sigma_q \, . 
\ee

For fixed forcing strength $q$, and a distribution of angle given by
equation (\ref{angleop}), the first moment of the distribution becomes 
\be
\langle g_k \rangle = - {1 \over \pi} 
\left( {\sin \aint \over \aint} - \cos \aint \right) - {q \over 2} 
\left( 1 + {\sin \aint \over \aint} \right) \, , 
\ee 
and the second moment is given by 
$$\
\langle g_k^2 \rangle = {1 \over \pi^2} 
\left[ {\aint^2 \over 6} - {\cos (2\aint) \over 4\aint} - 
{2\aint^2 -1 \over 8\aint} \sin(2\aint) \right] 
+ {q^2 \over 4} \left[ {3 \over 2} + 
{2 \sin \aint \over \aint} + {\sin (2\aint) \over 4\aint} \right]
$$
\be   
+ {q \over \pi} \left[ - \cos \aint - {1 \over 4} \cos (2\aint) 
+ {\sin \aint \over \aint} + {\sin (2\aint) \over 8\aint} \right]  \, . 
\ee
These expressions are somewhat cumbersome; if we take the angular
interval to be $\aint$ = $2 \pi m$, where $m$ is an integer as before,
all of the sine terms vanish and the moments simplify to the forms 
\be
\langle g_k \rangle = {1 \over \pi} - {q \over 2} \qquad {\rm and} \qquad 
\langle g_k^2 \rangle = {1 \over \pi^2} 
\left( {\aint^2 \over 6} - {1 \over 4\aint} \right) 
+ {3 q^2 \over 8} - {5 q \over 4 \pi}    \, . 
\ee
In this case, the variance is given by 
\be
\sigma_g^2 = {1 \over \pi^2} \left( {\aint^2 \over 6} - {1 \over 4\aint} - 1 
\right) + {q^2 \over 8} - {q \over 4 \pi} \, . 
\ee
In the limit $\aint \to \infty$, the width of the distribution does 
not converge, but rather diverges linearly so that 
\be
\lim_{\aint \to \infty} \sigma_g \sim {\sqrt{6} \over \pi 6 } \, \aint \, . 
\ee

As before, we can simultaneously include the distributions of angle
and forcing parameter, provided that the variables are sampled in an
independent manner. For the case where the angular interval is taken 
to be $\aint = 2 \pi m$, the moments reduce to the forms
\be
\langle g_k \rangle = {1 \over \pi} - {1 \over 2} \langle q \rangle 
\qquad {\rm and} \qquad \langle g_k^2 \rangle = {1 \over \pi^2} 
\left( {\aint^2 \over 6} - {1 \over 4\aint} \right) + {3 \over 8} 
\langle q^2 \rangle - {5 \over 4\pi} \langle q \rangle^2  \, . 
\ee
The variance for this case is given by 
\be
\sigma_g^2 = {1 \over \pi^2} \left( {\aint^2 \over 6} - 
{1 \over 4\aint} - 1 \right) + {1 \over 4} \left\{ 
{3 \over 2} \langle q^2 \rangle - \langle q \rangle^2 \right\} 
- {1 \over 4 \pi} \langle q \rangle \, .  
\ee
As a way to compare with the results found for the moments of 
the $h_k$ distribution, we again consider the limit where both 
$\langle q \rangle$ and $\aint$ are large. In this case, the 
variance of the $g_k$ has the order 
\be
\sigma_g^2 = {\cal O} (\aint^2) + {\cal O} (\sigma_q^2) \, . 
\ee
Comparing this result with equation (\ref{orderhsig}), we 
find that when the angular interval $\aint$ is not too large, 
the variance of the $h_k$ and that of the $g_k$ are of the same 
order. When the interval $\aint$ is large, however, the variance 
of the $g_k$ dominates.  

Next we consider the limit of large forcing parameters $q_k$. In this
limit, the leading order growth rate is given by Theorem 2.1, and the
corrections are determined by the variables $x_k$ and $\phi_k$
appearing in the matrix of equation (\ref{mbdefine}).  For the sake of
definiteness, we define the mean $q_0 = \langle q_k \rangle$, with
$q_0$ large, and allow the $q_k$ to have large variations about the
mean, but not so large that $q_k \to 0$. Under these conditions, one
can show that the variances of the variables $x_k$ and $\phi_k$ (from
equation [\ref{mbdefine}]) are of the order
\be
\sigma_x^2 = {\cal O} (q_0^{-2}) \qquad {\rm and} \qquad
\sigma_\phi^2 = {\cal O} (q_0^{-4}) \, . 
\label{sigorder} 
\ee 
As a result, the variance is dominated by the $x_k$ rather than by the
$\phi_k$. This result is consistent with the findings of Section II. 

\bigskip
\bigskip
\noindent
{\bf VI. CONCLUSIONS AND DISCUSSION} 
\bigskip

This paper has generalized and extended previous work concerning
Hill's equations (\ref{basic}) that contain random forcing parameters,
with a focus on the case of delta function barriers (equation
[\ref{qhatdelta}]).  In this formulation of the problem, both the
natural oscillation frequency $\af_k$ and the forcing strength $q_k$
can vary from cycle to cycle.  The development of the solutions to
Hill's equation, including the growth rates for instability, are given
by the general problem of matrix multiplication (equation
[\ref{mapzero}]), where the matrix elements are determined by the
principal solutions for a given cycle.  We have constructed the
principal solutions for individual cycles using Dirac delta functions
as the periodic barriers (equation [\ref{deltaprin}]).  This
construction allows us to explicitly show how both the matrix elements
$h_k$ and $g_k$, and the growth rates $\gamma$, depend on the
distributions of the original parameters $(\af_k, q_k)$ appearing in
Hill's equation (Section V).

In the limit of large forcing strength parameters $q_k$, the growth
rates approach the form $\gamma \sim \langle \log |q_k| \rangle$ (see
Theorem 2.1).  In this limit, large $q_k$ values often lead to large
values of the principal solutions at the end of the cycle; this
result, in turn, demonstrates (by construction) that the highly
unstable limit [2] can be realized using unremarkable values of the
parameters. In the opposite limit of small forcing parameters, the
growth rates approach the form $\gamma \sim \langle q_k^2 \rangle$
(Theorem 3.1).  Our results in this limit show how the fluctuations in
the Hill's equation parameters act to fill in the bands of stability
in the classic problem with fixed parameters (Figure \ref{fig:compare}).  
We have also found analytic results for the widths of the remaining
bands of stability (where the growth rates vanish). In this same limit
(small $q_k$), the Fokker-Planck equation provides an alternate
description of the dynamics (Section IV) and consistent estimates for
the growth rates.  Finally, we have constructed a iterative map
(Appendix C) that provides a heuristic argument for the general form
of the growth rates in the limits of both large and small forcing
strength $q_k$.

The results of this work can be used in a number of applications. For
example, one motivation for considering random Hill's equations was to
study orbital instabilities in extended mass distributions, such as
dark matter halos, galactic bulges, and young embedded star clusters
(see Appendix A). The results presented herein show when the orbits
are unstable and provide estimates for the corresponding growth rates.
These results, in turn, help explain the observed dynamical structures
in these astrophysical systems. Another important application involves
the reheating problem at the end of the inflationary epoch in the
early universe (see Appendix B). In this context, the introduction of
stochastic perturbations (e.g., due to quantum fluctuations) leads to
the disappearance of the bands of stability (see Figures
\ref{fig:deltaqk} and \ref{fig:compare}). As result, fluctuations
enhance the effectiveness of the reheating process. In addition to
these motivating examples, random Hill's equations arise in a wide
variety of other physical problems [2--5,7,14,15,19--21].

\newpage 
\bigskip 
\bigskip 
\bigskip 
\noindent
{\bf ACKNOWLEDGMENTS} 
\bigskip 

The work of FCA and AMB is jointly supported by NSF Grant DMS-0806756
from the Division of Applied Mathematics, and by the University of
Michigan through the Michigan Center for Theoretical Physics. AMB is
also supported by the NSF through grants CMS-0408542 and DMS-604307.
FCA is also supported by NASA through the Origins of Solar Systems
Program via grant NNX07AP17G.

\bigskip 
\bigskip 
\renewcommand{\theequation}{A\arabic{equation}}
\setcounter{equation}{0}  

\noindent{\bf APPENDIX A: } 

\noindent{\bf RANDOM HILL'S EQUATION FROM ASTROPHYSICAL ORBITS} 
\bigskip 

One application of Hill's equation with random forcing terms involves
the study of an instability that affects orbits in extended mass
distributions, such as dark matter halos [4]. In this setting, the
density profile $\rho(\varpi)$ of the halo has the general form given by
\be 
\rho (\varpi) = \, \rho_0 \, {F(\varpi) \over \varpi} \, , 
\label{rhogeneral} 
\ee 
where $\rho_0$ is a density scale and the variable $\varpi$ is written
in terms of the usual $(x,y,z)$ coordinates through the relation
\be
\varpi^2 = {x^2 \over a^2} + {y^2 \over b^2} + {z^2 \over c^2} \, , 
\label{mdef} 
\ee 
where $a > b > c > 0$.  The density field is thus constant on
ellipsoids. The function $F(\varpi)$ is approaches unity as $\varpi
\to 0$ so that the density profile approaches the form $\rho \sim
1/\varpi$ in the inner limit. For this regime, one can find 
analytic forms for both the potential and the force terms [4].

Further, when an orbit begins in any of the three principal planes,
the motion can be unstable to perturbations in the perpendicular
direction. Consider an orbit initially confined to the $x-z$ plane,
with a small perturbation in the perpendicular $\hat y$ direction.  
In the limit $|y| \ll 1$, the equation of motion for the $y$-coordinate 
takes the form 
\be 
{d^2 y \over dt^2} + \omega_y^2 y = 0 \qquad {\rm where} \qquad 
\omega_y^2 = { 4/b \over \sqrt{c^2 x^2 + a^2 z^2} + b \sqrt{x^2 + z^2} } \ . 
\label{omegay} 
\ee 
In this setting, the time evolution of the coordinates $(x,z)$ is
determined by the original orbit. Since this orbital motion is nearly
periodic, the $[x(t),z(t)]$ dependence of the parameter $\omega_y^2$
provides a periodic forcing term. The orbit has a maximum extent
(outer turning points) which results in a minimum value for
$\omega_y^2$, which in turn defines the natural oscillation frequency
$\af_k$. The parameter $\omega_y^2$ defined above can thus be written 
in the form 
\be
\omega_y^2 = { 4/b \over \sqrt{c^2 x^2 + a^2 z^2} + b \sqrt{x^2 + z^2} } = 
\af_k + Q_k (t) \, , 
\label{expandomega} 
\ee
where the index $k$ counts the number of orbit crossings, and the
chaotic orbit in the original plane leads to different values of
$\af_k$ and $Q_k(t)$ for each crossing. The shape of the functions
$Q_k$ are nearly the same, however, so that one can write $Q_k (t) =
q_k \qhat(t)$, where the forcing strength parameters $q_k$ vary from
cycle to cycle. These forcing strengths $q_k$ are determined by the
inner turning points of the orbit (with appropriate weighting from the
axis parameters $[a,b,c]$).  Given the expansion of equation
(\ref{expandomega}), the equation of motion (\ref{omegay}) for the
perpendicular coordinate becomes a random Hill's equation, with the
form of equation (\ref{basic}), as studied herein.

\bigskip 
\bigskip 
\renewcommand{\theequation}{B\arabic{equation}}
\setcounter{equation}{0}  

\noindent {\bf APPENDIX B: } 

\noindent {\bf RANDOM HILL'S EQUATION FROM REHEATING IN INFLATION}
\bigskip

In the inflationary universe paradigm [12], the accelerated expansion
of the universe is (usually) driven by the vacuum energy associated
with a scalar field $\varphi$ (often called the inflaton).  During the
phase of accelerated expansion, the energy density of the universe
itself decreases exponentially and the cosmos becomes relentlessly
empty. This epoch is thought to take place when the universe is
extremely young, with typical time scales of $\sim 10^{-36}$ sec. 

In order for the inflationary epoch to solve the cosmological issues
it was designed to alleviate, the end of inflation must involve a
mechanism to fill the universe with energy (e.g., see the review in 
Ref. [17]). This process is called reheating. During the epoch of
reheating, the equation of motion for the inflaton field displays
oscillatory behavior about the minimum of its potential. Further, in
order for the universe to become filled with energy (reheat), the
inflaton field $\varphi$ must couple to matter or radiation fields.
One simple type of interaction that is often considered uses an
coupling term in the Lagrangian of the form
\be
{\cal L}_{\rm int} = g \varphi \chi^2 \, , 
\ee
where $\chi$ is a second scalar field that represents matter
(radiation) and where the coupling constant $g$ sets the strength of
the interaction. The field $\chi$ is generally expanded in terms of
its Fourier modes $\chi_k$ since these quantities evolve
independently.  The resulting equation of motion for the matter field
modes $\chi_k$ takes the form
\be
{d^2 \chi_k \over dt^2} + 
\left[ \omega_k^2 + p(t) + q(t) \right] \chi_k = 0 \, , 
\ee
where $p(t)$ is a periodic function (given by the oscillatory behavior
of the inflaton field) and $q(t)$ is a noise term that provides
perturbations to the driving term $p(t)$ [21]. Note that the index $k$
refers here to the Fourier mode, although the forcing terms do vary
from cycle to cycle. In the absence of fluctuations, the matter field
modes $\chi_k$ thus obey a type of Hill's equation, which is subject
to parametric instability [15,16]. The noise perturbations convert the 
equation into a random Hill's equation [15,16,21], of the type studied 
herein. This type of equation was solved numerically using WKB methods 
[16], thereby finding the relevant physical solutions; nonetheless, the 
formulation of this paper can be applied to this class of reheating 
problems, and more general results can be obtained.  

\bigskip 
\bigskip 
\renewcommand{\theequation}{C\arabic{equation}}
\setcounter{equation}{0}  

\noindent{\bf APPENDIX C: AN ITERATIVE MAP} 
\bigskip 

As shown in the text, the growth rates for Hill's equation depend on
the forcing strength $q_k$ according to $\gamma \sim \langle q_k^2
\rangle$ in the limit of small symmetric $q_k$, and $\gamma \sim
\langle \log |q_k| \rangle$ in the limit of large $q_k$. These results
hold both for the particular case of delta function barriers
(considered here), and for the general problem [3]. In this
Appendix, we construct a heuristic argument that reproduces these
forms for the growth rate in the two limits. This treatment is highly
approximate, by design, but allows for a simple interpretation of our
previously obtained results.

Given the form of Hill's equation in the delta function limit, 
the jump condition across the barrier takes the form
\be 
{dy \over dt} \Big|_{+} = {dy \over dt} \Big|_{-} - 
q_k y  \, , 
\ee
where all of the functions are evaluated at the barrier. If we define
$V \equiv dy/dt$, and relabel the functions with an index $k+1$ on the
far side of the barrier, and an index $k$ on the near side, we obtain
an iterative map of the form
\be
V_{k+1} = V_k \left[ 1 - q_k {y_k \over V_k} \right] \, =
V_0 \prod_{k=1}^N \left[ 1 - q_k {y_k \over V_k} \right] \, , 
\ee
where we have continued the iteration back to the initial step to
obtain the second equality.  The growth rate $\gamma$ for this map 
can then be defined according to
\be
\gamma = \lim_{N \to \infty} {1 \over N} 
\sum_{k=1}^N \log \left| 1 - q_k {y_k \over V_k} \right| \, . 
\ee 
Given the form of Hill's equation away from the delta function
barrier, the solutions are oscillatory with frequency $\sqrt{\af_k}$, 
so that the function $y_k$ and the velocity are related via
\be 
{y_k \over V_k} = {1 \over \sqrt{\af_k}} F (\sqrt{\af_k} \, t) \, , 
\label{ratioest} 
\ee
where the function $F$ depends on the angle $\sqrt{\af_k} \, t$. 
If we use the ansatz implied by equation (\ref{ratioest}), 
the growth rate takes the form 
\be
\gamma = \lim_{N \to \infty} {1 \over N} 
\sum_{k=1}^N \log \left| 1 - {q_k \over \sqrt{\af_k}} 
F \right| \, . 
\ee 

In the limit of large forcing strength $q_k \gg 1$, the 
growth rate reduces to the form 
\be
\gamma = \lim_{N \to \infty} {1 \over N} \sum_{k=1}^N 
\log \left| {q_k \over \sqrt{\af_k}} F \right| \sim 
\left\langle \log \left| {q_k \over \sqrt{\af_k}} \right| 
\right\rangle \sim \left\langle \log \left| h_k \right| 
\right\rangle \, , 
\ee 
where we have ignored the function $F$ of the angle in obtaining the
final approximate forms. This argument reduces the problem to a single
(approximate) iterated jump condition, but still reproduces the proper
dependence of the growth rate for the highly unstable limit ($\gamma
\sim \langle \log \left| h_k \right| \rangle$).  We note that ignoring
the angular function is not valid when $F \to 0$. As shown above, this
problem allows for narrow bands of stability where the growth rate can
vanish even when the forcing strength is large. The presence of stable 
behavior ($\gamma \to 0$) can thus be accounted for through this
heuristic argument (by allowing $F \to 0$).

In the opposite limit of small forcing strength $|q_k| \ll 1$,
we can expand the logarithmic function in the expression for 
the growth rate to obtain
\be
\gamma = \lim_{N \to \infty} {1 \over N} 
\sum_{k=1}^N \left| {q_k \over \sqrt{\af_k}} F + 
{q_k^2 \over 2 \af_k} F^2 \right| \, . 
\ee 
For symmetric fluctuations, the first term vanishes in the limit, 
so that the quadratic term provides the leading order contribution 
to the growth rate. Ignoring the angular function $F$ as above, the 
growth rate becomes 
\be
\gamma \sim \left\langle {q_k^2 / \af_k} \right\rangle \, . 
\ee
In this case, the iterated jump condition argument reproduces the
proper dependence of the growth rate for the limit of symmetric and
weak forcing (compare with Theorem 2.1). 

\bigskip 
\bigskip 
\newpage 
\noindent 
{\bf REFERENCES} 
\bigskip 

\medskip\noindent
[1] Abramowitz, M., and Stegun, I. A., 
{\it Handbook of Mathematical Functions} (Dover, New York, 1970).

\medskip\noindent
[2] Adams, F. C.,  and Bloch, A. M., 
``Hill's Equation with random forcing terms,''
SIAM J. Appl. Math. {\bf 68}, pp. 947 -- 980 (2008).  

\medskip\noindent
[3] Adams, F. C., and Bloch, A. M., 
``Hill's Equation with random forcing parameters:
General treatment including marginally stable cases,''
submitted to J. Stat. Phys. (2009). 

\medskip\noindent
[4] Adams, F. C., Bloch, A. M., Butler,  S. C., Druce, J. M., and Ketchum, J. A., 
``Orbits and instabilities in a triaxial cusp potential,'' Astrophys. J., 
{\bf 670}, pp. 1027 -- 1047 (2007). 

\medskip\noindent
[5] Anderson, P. W., ``Absence of diffusion in certain random lattices,''
Physical Review {\bf 109}, pp. 1492 -- 1505 (1958). 

\medskip\noindent
[6] Binney, J. and Tremaine, S., {\it Galactic Dynamics}, 
(Princeton Univ. Press, Princeton, 1987).

\medskip\noindent
[7] Cambronero, S., Rider, B., and Ramer{\'i}z, J., ``On the shape of
the ground state eigenvalue density of a random Hill's equation,'' 
Comm. Pure Appl. Math. {\bf 59}, pp. 935 -- 976 (2006).   

\medskip\noindent
[8] Cohen, J. E., and Newman, C. M., ``The stability of large random
matrices and their products,'' Annals of Prob. {\bf 12}, pp. 283 -- 310 (1984). 

\medskip\noindent
[9] Doering, C. R., and Gradoua, J. C., ``Resonant activation 
over a fluctuating barrier,'' Phys. Rev. Lett. {\bf 69},
pp. 2318 -- 2321  (1992).

\medskip\noindent
[10] Furstenberg, H., ``Noncommuting random products,'' Trans. Amer. 
Math. Soc. {\bf 108}, pp. 377 -- 428 (1963). 

\medskip\noindent
[11] Furstenberg, H., and Kesten, H., ``Products of random matrices,'' 
Ann. Math. Statist. {\bf 31}, pp. 457 -- 469 (1960). 

\medskip\noindent
[12] Guth, A. H., ``Inflationary Universe: A possible solution to the
horizon and flatness problems,'' Phys. Rev. D {\bf 23}, pp. 347 -- 356 (1981).

\medskip\noindent
[13] Hill, G. W., ``On the part of the motion of the lunar perigee
which is a function of the mean motions of the Sun and Moon,''
Acta. Math. {\bf 8}, pp. 1 -- 36 (1886). 

\medskip\noindent
[14] Ishii, K., ``Localization of eigenstates and transport phenomena
in one-dimensional disordered systems,'' Progress Theor. Phys. Suppl.
{\bf 45}, pp. 77 -- 119 (1973).

\medskip\noindent
[15] Kofman, L., Linde, A., and Starobinsky, A. A., 
``Reheating after Inflation,'' Phys. Rev. Lett., 
{\bf 73}, pp. 3195 -- 3198 (1994).

\medskip\noindent
[16] Kofman, L., Linde, A., and Starobinsky, A. A., 
``Towards the theory of reheating after Inflation,'' 
Phys. Rev. D {\bf 56}, pp. 3258 -- 3295 (1997). 

\medskip\noindent
[17] Kolb, E. W., and Turner, M. S., {\it The Early Universe}, 
(Addison-Wesley, Reading MA, 1990). 

\medskip\noindent
[18] Lima, R., and Rahibe, M.,
``Exact Lyapunov exponent for infinite products of random matrices,'' 
J. Phys. A. Math. Gen. {\bf 27}, pp. 3427 -- 3437 (1994). 

\medskip\noindent
[19] Magnus, W., and Winkler, S., {\it Hill's Equation}, (Wiley, New York, 1966). 

\medskip\noindent
[20] Pastur, L., and Figotin, A., 
{\it Spectra of Random and Almost-Periodic Operators}, a Series of
Comprehensive Studies in Mathematics, (Springer-Verlag, Berlin, 1991).

\medskip\noindent
[21] Zanchin, V., Maia, A., Craig, W., and Brandenberger, R., 
``Reheating in the presence of noise,'' Phys. Rev. D. {\bf 57},  
pp. 4651 -- 4662 (1998). 

\end{document}